\documentclass[final,11pt,times]{elsarticle}

\usepackage{amssymb}
\usepackage{booktabs}
\usepackage{geometry}
 \usepackage{listings}
 \usepackage{subcaption}
 \usepackage{graphicx}

\usepackage{fancyhdr}
\journal{CIE53 Proceedings}

\usepackage[colorlinks=true, pdfauthor={author}]{hyperref}

\begin{document}

\begin{frontmatter}


\title{Fault Detection and Explainable Classification in Automotive HIL Validation via Denoising Autoencoders and In-Context Large Language Models}


\author{Mohammad Abboush \corref{cor}}
\ead{mohammad.abboush@tu-clausthal.de}
\author{Hamza Ouarrad}
\ead{hamza.ouarrad@tu-clausthal.de }
\author{Andreas Rausch }
\cortext[cor]{Corresponding author}
\ead{andreas.rausch@tu-clausthal.de}
\address{Clausthal University of Technology, Arnold-Sommerfeld-Straße 1, 38678 Clausthal, Germany}

\begin{abstract}
Validating automotive software systems produces large multivariate test recordings that are still examined through effort-intensive manual review and rule-based evaluation, which detects faults beyond predefined rules poorly. Machine and deep learning have advanced fault diagnosis, yet most supervised models require large labelled datasets, generalise poorly to unseen conditions, and offer little insight into their decisions. We propose a generalisable and explainable two-phase framework for fault detection and classification during real-time validation. A denoising autoencoder trained only on healthy signals first flags abnormal behaviour through reconstruction-error analysis, removing the need for fault labels. Each abnormal window is then encoded as compact textual statistical evidence relative to a time-aligned healthy reference and classified by a frozen large language model under zero-shot and few-shot prompting, returning the predicted class, ranked alternatives, a confidence value, the fault location, and a short evidence-based explanation. Eight open-source models are evaluated across two powertrains and three driving regimes. The detector attains average F1-scores of 0.97 across powertrains and 0.98 across regimes, with average mean error below 0.03. Zero-shot prompting proves insufficient (best 0.519 F1-score), whereas few-shot prompting reaches perfect discrimination under stable regimes, showing that prompting strategy, rather than parameter count, governs classification quality: a nine-billion-parameter model surpasses every zero-shot medium and large model. Mistral Small 24B is adopted as the main pipeline model for its balance of accuracy, class-balanced reliability, calibration, and inference cost, giving engineers interpretable diagnostic reports and more efficient validation.
\end{abstract}


\end{frontmatter}
\thispagestyle{fancy}

\section{Introduction}

In the automotive domain, the integration of software-controlled functions into
decision-making has enabled higher autonomy, intelligence, and connectivity
\cite{kukkala2018advanced} and improved operational efficiency and driving comfort
\cite{golias2002classification}, but it has also markedly increased system
complexity. Modern automotive software systems (ASSs) combine heterogeneous
hardware, sensors, actuators, electronic control units (ECUs), and real-time
communication networks \cite{d2017systems}, with vehicles containing more than 120
ECUs that exchange over two million messages per minute across multiple buses
\cite{ebert2017automotive}; as this complexity grows, so does the probability of
failures.

Such failures may stem from software errors, sensor wear, data-stream
inconsistencies, or unexpected interactions with the environment
\cite{parhizkar2023degradation}, with effects ranging from minor degradation to
safety-critical incidents; a NASA report, for example,
linked unintended acceleration in Toyota vehicles to faults in the accelerator
pedal position system \cite{koopman2014case}.

To ensure functional safety, ASS development must comply with ISO~26262 throughout
the lifecycle \cite{url}. Within the V-model, later-phase validation typically uses
public-road test drives that assess safety goals under realistic conditions
\cite{klitzke2019real} and require broad coverage of normal and edge-case scenarios
across control units, sensors, actuators, and communication networks
\cite{samuel2002automotive}. As an effective alternative, real-time
hardware-in-the-loop (HIL) simulations integrate real ECUs and physical subsystems
into a controlled environment, enabling safe, flexible, and repeatable validation
\cite{abboush2024virtual}. Owing to the complexity of the system under test (SUT),
both approaches generate extensive datasets marked by multimodality, nonlinearity,
heterogeneity, high dimensionality, redundancy, data imbalance, and the rarity of
critical events.

These logs are still analysed mainly through manual review, expert assessment, and
rule-based evaluation against predefined expectations and safety criteria
\cite{theissler2013detecting}. As the volume and diversity of data grow, millions
of measurements and control logs must be examined to determine root causes and
trace fault propagation \cite{szalay2019proof}, and the effort and cost rise with
the number of scenarios and configurations. Conventional
approaches also cannot detect novel fault types absent from predefined rules or
historical data, so intelligent and automated methods are required.

Data-driven fault detection and diagnosis (FDD) overcomes the reliance of
model-based and knowledge-based 
methods on accurate models and expert knowledge. Machine
learning (ML) and deep learning (DL) have shown strong performance in automotive
fault classification \cite{said2025deep}, yet they generalise poorly to unseen
conditions and configurations \cite{baccari2024anomaly}, assume stationary
distributions, and require retraining from scratch when new faults emerge, which
limits efficiency and scalability in safety-critical settings
\cite{zhang2025survey}; they also frequently lack interpretability. Whereas
supervised DL needs large labelled datasets covering many fault classes
\cite{lei2025research}, unsupervised reconstruction-based techniques detect
abnormal behaviour without labels. Specifically, denoising autoencoders (DAEs) learn
representative latent features of high-dimensional temporal signals under noise
\cite{vincent2008extracting}, so reconstruction-error analysis improves robustness
against noise.

In parallel, large language models (LLMs) such as GPT, Llama, Qwen, and Phi excel
at logical reasoning and processing heterogeneous information
\cite{zheng2024empirical}, and recent work applies them to fault diagnosis by
converting time-series data into text. Owing to self-supervised pre-training on
cross-domain corpora, they show strong in-context generalisation and can be adapted
through prompting alone, without parameter updates. Their generalisation under
realistic automotive validation, however, remains underexplored, and the effects of
model selection, input representation, and prompting strategy on performance and
explainability are not yet well understood; prompting that incorporates
fault-related features, system behaviour, and observations can support
context-aware reasoning, while frozen parameters reduce adaptation cost and memory
at deployment.

This study therefore investigates LLMs for classifying sensor-related faults during
ASS development, with particular attention to the generalisation aspect with few-shot examples. The framework represents sensor data and metadata as compact, structured textual evidence and classifies through prompting, without task-specific parameter updates; open-source LLMs of different sizes are evaluated under zero-shot and few-shot prompting across model families and scales. Analysis proceeds in two phases: a DAE detects faults, and a frozen LLM classifies them and provides explanations, which reduces runtime overhead while preserving detailed analysis of abnormal conditions and supporting
test engineers with interpretable explanations. To the best of our knowledge, this
is the first study to apply LLMs to digital test-drive datasets generated from HIL
simulations, considering faults that propagate through physically and functionally
coupled signals during the real-time validation phase of ASSs.

The main contributions of this work are summarised as follows:

\begin{itemize}
    \item A two-phase framework decoupling label-free detection (DAE on healthy
    signals) from interpretable classification (frozen LLM over compact,
    healthy-referenced evidence), cutting runtime overhead while retaining detailed
    analysis of abnormal windows.

    \item The first application of LLMs to real-time HIL test-drive data, evaluated within a leakage-controlled temporal protocol on two powertrains (gasoline-engine and electric-vehicle) and three driving regimes.

    \item An evidence representation separating direct from propagated deviations,
    constraining the model to a structured output (class, ranked alternatives,
    confidence, fault location, explanation) for root-cause reasoning and traceable
    justification.

    \item A systematic evaluation of eight open-source LLMs under zero-shot and
    few-shot prompting, quantifying the effects of input representation, prompting
    strategy, and model scale on classification, ranking, calibration, and runtime.

    \item Evidence that prompting strategy, not parameter count, governs
    classification quality, with Mistral Small~24B chosen for its balance of
    accuracy, MCC, calibration, and inference cost.

\end{itemize}

The remainder of the paper is organised as follows. Section~2 reviews the state of
the art, Section~3 presents the methodology, Section~4 describes the experimental
setup and case study, Section~5 reports and discusses the results with a
comparative analysis, and Section~6 concludes and outlines future work.

\section{Related work}

This section aims to shed light on recent advancements in the field of data-driven FDD for ASSs, highlighting the key contributions and the limitations imposed by methodological frameworks.

Data-driven ML and DL have been applied across phases of the software development life cycle (SDLC) \cite{shafiq2021literature}. This is due to the noteworthy accomplishments achieved compared with conventional methodologies. In automotive system validation, many ML and DL methods have been proposed for automated fault identification, classification, isolation, and characterisation from historical datasets \cite{theissler2021predictive}. For instance, Safavi et al. in \cite{safavi2021multi} developed 1D-CNN and multi-class DNN architectures to detect and identify erratic, hard-over, spike, and drift faults as a sensor-related fault during real test drives. In a similar context, an ensemble classifier-based approach was presented in \cite{theissler2017detecting} with the aim of detecting the anomaly under various driving conditions. This approach supports the analysis process of real-world vehicle test records, achieving an average F2 score of 80\%. In a similar manner, Weibull-calibrated one-class support vector machines (SVMs) were employed in conjunction with Bayesian filtering to identify unknown faults in \cite{jung2020data}, resulting in an accuracy of 85\% for seven distinct categories of engine-related faults. Focusing on electric vehicles (EVs), an LSTM-based framework for diagnosing short-circuit and open-circuit faults was proposed, achieving an average accuracy of 97.05\% \cite{kaplan2021fault}.

In the early phase of ASSs development, investigating the applicability of ML and DL methods in validation processes remains limited. Nevertheless, several studies have demonstrated the effectiveness of integrating intelligent diagnostic-based ML approaches into real-time validation processes with HIL environments. For instance, \cite{raveendran2020brake} investigated fault diagnosis methods for vehicle braking systems based on random forest techniques, while \cite{pietrowski2024fault} proposed an algorithm for diagnosis fault-related EPS based on real-time HIL simulation data. Furthermore, a hybrid DL-based approach for detecting and identifying sensor-related faults during HIL testing has been proposed by \cite{abboush2023intelligent}, utilising CNN-LSTM and DAE, with the aim of improving robustness under noisy conditions.

Concurrently, LLMs, as pre-trained models on extensive cross-domain datasets, have demonstrated superiority in logical reasoning and the processing of diverse information in comparison to DL models \cite{xu2025large}. Consequently, the employment of LLMs within the domain of software testing has prompted interest from researchers within both academic and industrial contexts \cite{wang2024software}. The proposed methods are predicated on the premise of converting time-series data into textual representations, followed by an efficient fine-tuning process. 

In the context of optimising the testing of automated vehicles, a sophisticated LLM-powered intelligent agent has been developed for several purposes, including test case generation,  HIL test automation, and vehicle API verification. For example, in \cite{danso2025automated}, an LLM-based pipeline was proposed for the automatic generation of scenarios for testing the functionality of autonomous vehicles.  Similarly, in industrial HIL environments, \cite{feng2025smarter} proposed the HIL-GPT framework, which aims to support the testing process within HIL simulation. However, the proposed framework is limited to integrating domain-adapted LLMs with semantic retrieval via vector indexing for scalable test case and requirement retrieval without focusing on test record analysis.

In addition, LLMs have emerged as a promising direction in industrial and automotive fault detection and diagnosis. The rationale behind this phenomenon lies in their capacity to process heterogeneous data and to provide explanations based on inferences. In the field of software engineering, Zheng et al. in \cite{zheng2024empirical} have proposed one of the earliest LLM-based diagnostic frameworks, which uses textual representations of system state data. Subsequent studies have proposed multimodal and knowledge-driven LLM frameworks for industrial monitoring and fault diagnosis \cite{lin2025fd}. Most existing approaches are restricted to single-machine datasets and address neither concurrent faults nor generalisation across conditions.  
Despite the noteworthy accomplishments of these studies in terms of fault diagnostic performance, they are encumbered by limitations with regard to the generalisability across a range of operating conditions and configurations in the automotive domain. Furthermore, there is a need for more in-depth investigation into its applicability in safety-critical environments in the domains of real-time HIL validation, considering the ISO 26262 standard. A review of current studies reveals a paucity of a unifying framework capable of combining fault detection and classification, explainable and generalizable diagnosis for real-time validation in the automotive sector within HIL-based environments across various conditions.

\section{Methodology}
\label{Methodology}

\begin{figure*}[t]
\centering
\includegraphics[width=\textwidth]{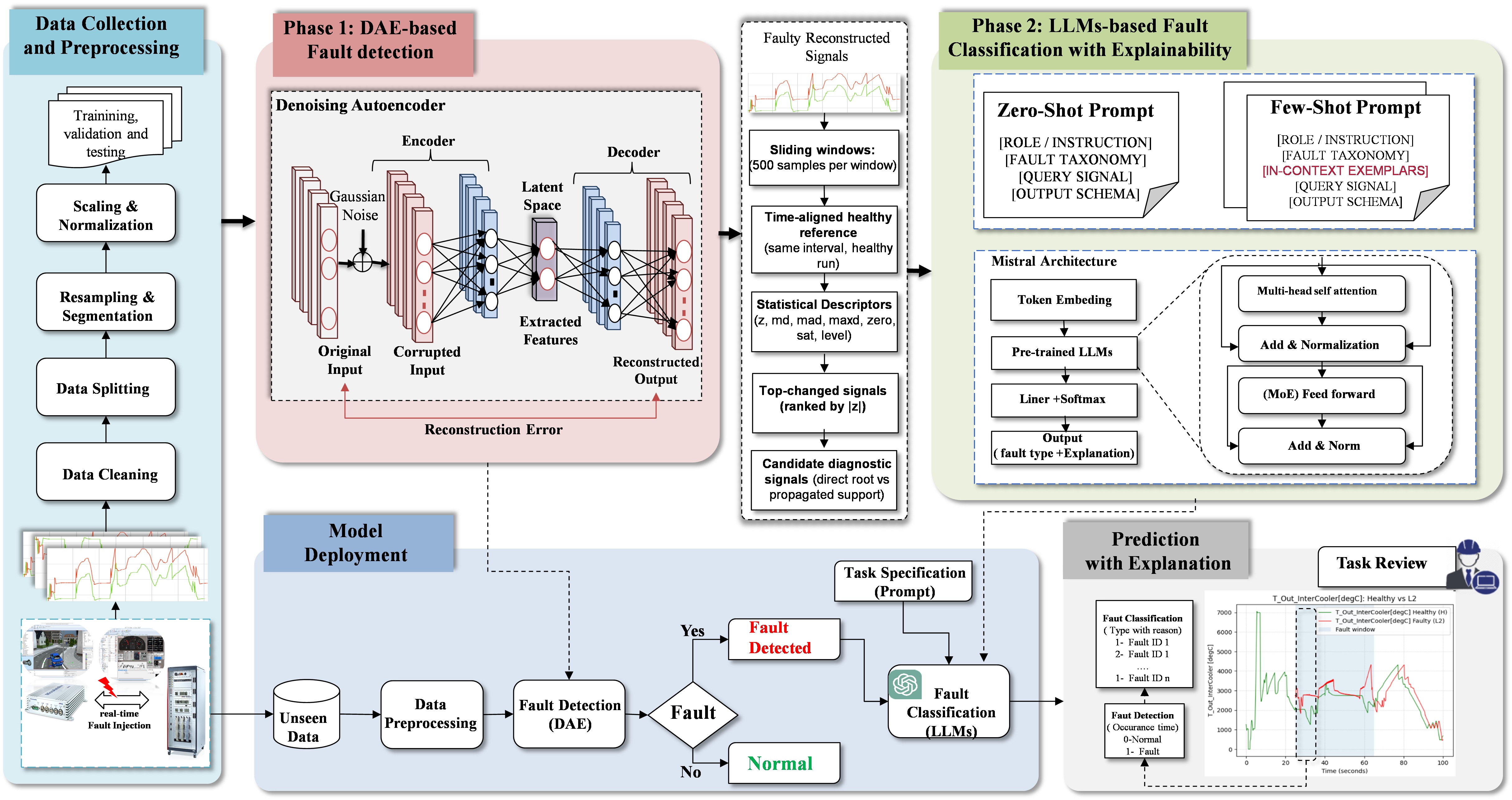}
\caption{Proposed two-stage framework for fault diagnosis in automotive HIL systems based on DAE and LLMs.}
\label{fig:framework}
\end{figure*}

The proposed methodology for fault analysis in automotive HIL validation is illustrated in Figure~\ref{fig:framework}. It consists of two diagnostic phases. In Phase~1, a DAE-based detector identifies deviations from nominal behaviour. In Phase~2, a frozen LLM classifies the detected fault window and returns an interpretable diagnostic output. This separation reflects the different nature of the two tasks. Fault detection requires sensitivity to deviations from healthy behaviour, whereas fault classification requires reasoning over heterogeneous evidence and candidate fault locations. The LLM parameters remain fixed throughout the experiments; task adaptation is achieved only through zero-shot and few-shot prompting.

\subsection{Data collection and preprocessing}
\label{sec:data}

Data acquisition takes place on a real-time HIL simulation platform in which faults are injected into running test scenarios. For every scenario, both a healthy execution and the corresponding faulty executions are recorded, so that each faulty measurement can later be related to the nominal behaviour of the same driving profile. The recorded multivariate signals cover driver inputs, vehicle dynamics, and powertrain variables.

The raw recordings are prepared in several consecutive steps. Data cleaning removes inconsistent samples and irrelevant information, scaling and normalisation compensate for the heterogeneous physical ranges of the system variables, and resampling and segmentation partition the signals into analysis segments. The prepared data are then divided into training, validation, and testing subsets, with only the fault-free portion used for the development of the detection model. Gaussian noise is added to the training inputs so that the autoencoder learns to denoise and reconstruct the underlying signal structure instead of memorising individual samples.

\subsection{Phase 1: DAE-based fault detection}
\label{sec:phase1}

The first phase deploys a denoising autoencoder (DAE) developed in \cite{abboush2025intelligent} for fault detection. The DAE is selected for its established effectiveness in industrial and automotive monitoring, where it extracts representative latent features under noisy conditions. As shown in Figure 1, the architecture comprises an input layer, a corrupted-input layer, encoder layers, a latent representation, decoder layers, and an output reconstruction layer. To learn stable representations, Gaussian noise is added to the healthy signals during training, and the reconstruction loss between the healthy input and its reconstruction is minimised.

The DAE is trained on healthy signals only, so the network learns the intrinsic characteristics of normal behaviour. At deployment, samples that deviate markedly from the learned representation yield higher reconstruction errors, signalling a fault, with the decision made by a predefined anomaly threshold. Unlike supervised learning, this threshold-based rule identifies abnormal behaviour without explicit fault labels, which is advantageous in automotive validation, where complete fault annotations are difficult to obtain.

The latent representation also aids detection by compressing high-dimensional temporal signals into fault-sensitive features while suppressing irrelevant variation and stochastic disturbance. Windows flagged as abnormal are then forwarded to the second phase, where compact statistical evidence is extracted relative to a time-aligned healthy reference for semantic interpretation and classification.

\subsection{Evidence extraction from HIL time-series windows}
\label{sec:evidence}

Raw time-series data are converted into compact diagnostic evidence, since long numerical sequences are inefficient for LLM processing. Analysis is restricted to the 15--45~s fault-effect interval using 5~s windows with a 1~s step size. Each faulty window is compared with the time-aligned healthy reference, and temporal leakage is prevented by using 15--25~s for few-shot examples, 25--30~s as an embargo region, and 30--45~s for testing.

For each signal, statistical descriptors relative to the healthy reference are computed and represented using compact keys (\texttt{sig}, \texttt{z}, \texttt{md}, \texttt{mad}, \texttt{maxd}, \texttt{zero}, \texttt{sat}, \texttt{level}). The prompt includes the top changed signals and candidate-specific evidence that separates direct root-signal evidence from propagated effects, encouraging root-cause reasoning rather than reliance on downstream deviations. The few-shot exemplars and the query windows originate from the same fault-injection runs but from disjoint, embargo-separated time segments. The protocol therefore measures adaptation under a temporal holdout rather than transfer across unseen runs or powertrains; the latter is examined separately in Section 5 and discussed as a limitation.

\subsection{Phase 2: LLM-based fault classification with explainability}
\label{sec:phase2}

Fault classification is formulated as a constrained language-modelling task over the extracted evidence. The zero-shot prompt serves as a structured-evidence baseline and contains the task instruction, the candidate fault classes, the query-window evidence, and the required output schema. The few-shot prompt extends this setting by adding labelled examples from the leakage-safe few-shot region before the unlabeled query window. In both settings, the healthy class is used only as a reference baseline and is excluded as a valid prediction class.

The instruction template is fixed, but the query content is generated separately for each test window. In the few-shot setting, the examples contain their own compact statistical evidence and correct labels, whereas the query window contains only unlabeled evidence. The true class of the query window is not inserted into the prompt. The candidate direct-vs-support evidence is generated for all candidate classes, not only for the correct one, so the model must compare the evidence across competing fault hypotheses.

The prediction can be expressed as
\begin{equation}
\hat{y} = \arg\max_{y \in \mathcal{Y}} P_{\Theta}\!\left(y \mid I, \mathcal{D}_{k}, s_{q}\right),
\label{eq:icl}
\end{equation}
where $\mathcal{Y}$ is the set of candidate fault classes, $I$ is the fixed instruction and taxonomy, $s_q$ is the evidence summary of the query window, and $\mathcal{D}_{k}$ denotes the labelled in-context examples. The case $k=0$ corresponds to zero-shot prompting, while $k>0$ corresponds to few-shot prompting. Since $\Theta$ remains frozen, no parameter update or task-specific fine-tuning is performed. In the final pipeline, Mistral Small~24B is used as the main LLM because it provides the best balance between diagnostic accuracy, confidence reliability, and inference cost. The model follows a decoder-only Transformer architecture, where the serialised prompt is processed autoregressively through self-attention layers. This allows the model to relate the task instruction, few-shot examples, candidate classes, and query-window evidence within a single context. No architectural modification or parameter fine-tuning is applied; the diagnostic behaviour is controlled only through the structured prompt and the compact evidence representation.

The prompt combines natural-language task instructions with JSON-style structured evidence. The model output is constrained to the following JSON schema:

\begin{lstlisting}[basicstyle=\ttfamily\small,breaklines=true]

{
  "predicted_class": "one candidate class",
  "confidence": 0.0,
  "ranked_classes": ["best", "second"],
  "fault_locations": ["location"],
  "explanation": "short root-evidence explanation"
}
\end{lstlisting}

The predicted class is used for Top-1 evaluation, while the ranked list is used for Top-2 evaluation. The confidence value is retained for calibration analysis, and the explanation field provides a concise justification of the diagnostic decision. The explanation is not used as the primary classification target, but it supports interpretability by indicating which direct and supporting signals influenced the model's decision.

\subsection{Model Deployment}
\label{sec:deployment}

During deployment, unseen HIL recordings are processed by the same preprocessing pipeline and evaluated by the DAE. Windows whose reconstruction error remains below the threshold are labelled as normal, while abnormal windows are converted into compact evidence and submitted to the frozen LLM. The returned JSON object is then compiled into a diagnostic report containing the detected health state, the predicted fault type, the ranked alternatives, the confidence value, and the textual explanation. Since each prediction is linked to explicit signal evidence, the test engineer can verify the diagnosis on technical grounds and use the output to support root-cause analysis.

\section{Case study and Implementation}

\subsection{HIL case study systems}
\label{sec:hil_systems}

The case study uses two automotive HIL platforms: a dSPACE ASM gasoline-engine system and an electric-vehicle powertrain system. Both operate in closed-loop real-time environments where injected faults propagate through coupled vehicle and control signals, providing realistic data for fault classification and explanation. The gasoline-engine setup includes driver, vehicle, and engine variables, while the EV setup focuses on battery, DC-link, electric-machine, steering, and vehicle-speed signals. Together, they evaluate the framework across two propulsion architectures. Due to limited per-class samples, results are reported with 95\% confidence intervals obtained using bias-corrected and accelerated bootstrap resampling.

\subsection{Dataset description and Prompt Templates}
\label{sec:data_description}

The dataset originates from automated fault-injection experiments in the HIL configurations described above. Each recording is one experimental run under healthy or faulty conditions, comprising approximately 7,700 samples at a fixed 0.01 s interval. The gasoline-engine system records 15 signal columns and the electric-vehicle system 17. The injected fault becomes active during the main part of the run and is compared against a time-aligned healthy reference trajectory.
The evaluation covers three scenarios and ten single-fault classes. The non-urban scenario contains accelerator-pedal, brake-pedal, engine-speed, steering, and throttle faults; the highway scenario contains high-voltage-battery, rear electric-machine-speed, and steering faults; and the urban scenario contains accelerator-pedal-gain and engine-speed-noise faults.
Two prompt templates are evaluated, where \texttt{<...>} denotes query-dependent content from the corresponding sensor window. The zero-shot prompt contains only query-window evidence, whereas the few-shot prompt adds labelled examples from the leakage-safe region together with a compact direct-evidence representation separating root-signal from propagated support evidence. Both exemplars and the query window use compact statistical descriptors, with class labels provided only for the exemplars. Prompts are generated per test window from a fixed instruction template with window-specific metadata, top changed signals, and candidate evidence. Natural-language instructions are combined with JSON-style structured evidence, and the output is constrained to a valid JSON object for automatic parsing and evaluation.

\section{Results and Discussion}

\subsection{Fault detection performance of the proposed DAE}

The generalisation behaviour of the proposed denoising autoencoder (DAE) was
further assessed across two dissimilar powertrain configurations, with the fault
detection results reported in Table~\ref{table:caseS}. For the gasoline engine case
study, the model recorded a precision of 0.9409 and a notably high recall of
0.9921, giving an F1-score of 0.9685 and a mean error of 0.0315. In the electric
vehicle (EV) case study, a contrasting pattern emerged, as precision rose to 0.9967
while recall settled at 0.9568, yielding an F1-score of 0.9763 and a reduced mean
error of 0.0237. Jointly, the DAE delivered an average precision of 0.9688, recall
of 0.9745, and F1-score of 0.9724, with the mean error limited to 0.0276. Since each
F1-score remains above 0.96, a consistently high detection capability is preserved
across two markedly different powertrains. The two scenarios prove complementary,
the gasoline configuration favouring near-perfect recall and the EV configuration
near-perfect precision, indicating that the model accommodates the distinct signal
characteristics of each system effectively. Overall, the balanced average F1-score
exceeding 0.97, together with a mean error below 0.03, attests to the robustness of
the proposed approach and its capacity to generalise reliably for fault detection in
automotive applications.

\begin{table*}
\centering
\footnotesize
\caption{Fault detection performance of the proposed DAE model under various case studies. \label{table:caseS}}%
\begin{tabular}{@{}lllll@{}}
\toprule
\textbf{Driving Scenario}  & \textbf{Precision} & \textbf{Recall} & \textbf{F1-score} & \textbf{Mean Error} \\
\midrule
Gasoline engine case study & 0.9409 & 0.9921 & 0.9685 & 0.0315 \\
EV case study              & 0.9967 & 0.9568 & 0.9763 & 0.0237 \\
\midrule
\textbf{Average performance} & \textbf{0.9688} & \textbf{0.9745} & \textbf{0.9724} & \textbf{0.0276} \\
\bottomrule
\end{tabular}
\end{table*}

To evaluate the generalisation capability of the developed DAE, fault detection performance was assessed across urban, non-urban, and highway driving scenarios (Table~\ref{tab:driving_scenarios}). The urban scenario achieved a precision of 0.9663, recall of 0.9783, F1-score of 0.9723, and mean error of 0.0277. Stronger results were obtained under non-urban (F1-score 0.9908, mean error 0.0092) and highway conditions (F1-score 0.9902, mean error 0.0098). Across all scenarios, the DAE achieved an average precision of 0.9808, recall of 0.9882, F1-score of 0.9844, and mean error of 0.0156.

These results demonstrate robust and consistent fault detection despite substantial differences in driving conditions. All scenarios achieved an F1-score above 0.97, while recall remained consistently high. Although urban driving produced slightly lower performance due to more dynamic signal behaviour, the DAE maintained a strong balance between precision and recall, confirming its ability to generalise reliably across heterogeneous automotive operating regimes.

\begin{table*}

\centering
\footnotesize
\caption{Fault detection performance of the proposed DAE model under different driving scenarios.}
\label{tab:driving_scenarios}
\begin{tabular}{lcccc}
\toprule
\textbf{Driving Scenario} & \textbf{Precision} & \textbf{Recall} & \textbf{F1-Score} & \textbf{Mean Error} \\
\midrule
Urban scenario     & 0.9663 & 0.9783 & 0.9723 & 0.0277 \\
Non-urban scenario & 0.9866 & 0.9951 & 0.9908 & 0.0092 \\
Highway scenario   & 0.9894 & 0.9911 & 0.9902 & 0.0098 \\
\midrule
\textbf{Average performance} & \textbf{0.9808} & \textbf{0.9882} & \textbf{0.9844} & \textbf{0.0156} \\
\bottomrule
\end{tabular}
\end{table*}

\subsection{Zero-Shot Baseline}

Classification is assessed by Top-1 and Top-2 accuracy (the true class among the k highest-ranked outputs), with macro-averaged precision, recall, and F1 to weight classes equally. Class-balanced agreement is given by the Matthews correlation coefficient (MCC), and confidence quality by the Brier score and expected calibration error (ECE), both lower-is-better. For detection, mean error is the average per-window reconstruction-error residual after thresholding.

Table~\ref{tab:zero_shot_small_grouped} reports the zero-shot performance of five small language models on the gasoline-engine and EV scenarios. Gemma~2~9B is the strongest model in both settings, and the only one marked best across the two. On the gasoline-engine scenario, it attains 0.529 Top-1 accuracy, 0.519 F1, and 0.425 MCC, together with the best calibration (Brier 0.245, ECE 0.213). On the EV scenario, it again leads, with 0.452 Top-1 accuracy, the highest Top-2 accuracy of 0.786, an F1 of 0.358, and an MCC of 0.318. The remaining models perform considerably worse: Llama~3.1~8B, Mistral~7B, and Qwen2.5~7B cluster around 0.33–0.41 Top-1 accuracy, while Phi-3.5~Mini collapses towards trivial behaviour, reaching only 0.029 and 0.143 Top-1 accuracy with the poorest calibration (ECE up to 0.860).

Even so, the best zero-shot accuracy remains well below a usable level. The wide Top-1 to Top-2 gap (0.452 against 0.786 on EV) shows the correct class is often ranked highly but rarely selected first, and the low MCC values confirm weak class-balanced agreement. Structured evidence alone is therefore insufficient for robust zero-shot classification, which motivates the few-shot study that follows.

\begin{table}[t]
\centering
\footnotesize
\caption{Zero-shot performance of small LLMs under various case studies. The best model in each scenario is highlighted in bold and marked with \checkmark.}
\label{tab:zero_shot_small_grouped}
\resizebox{\textwidth}{!}{%
\begin{tabular}{llcccccccc}
\toprule
Scenario & Model & Top-1 Acc & Top-2 Acc. & F1 & Precision & Recall & MCC & Brier $\downarrow$ & ECE $\downarrow$ \\
\midrule
Gasoline Engine & Llama 3.1 8B & 0.371 & 0.657 & 0.319 & 0.356 & 0.371 & 0.236 & 0.341 & 0.364 \\
 & Mistral 7B & 0.386 & 0.514 & 0.349 & 0.408 & 0.386 & 0.276 & 0.286 & 0.329 \\
 & Phi-3.5 Mini & 0.029 & 0.029 & 0.050 & 0.200 & 0.029 & 0.108 & 0.772 & 0.860 \\
 & Qwen2.5 7B & 0.357 & 0.543 & 0.322 & 0.545 & 0.357 & 0.249 & 0.337 & 0.326 \\
 & \textbf{Gemma 2 9B} \checkmark & \textbf{0.529} & \textbf{0.643} & \textbf{0.519} & \textbf{0.569} & \textbf{0.529} & \textbf{0.425} & \textbf{0.245} & \textbf{0.213} \\
\midrule
EV  & Llama 3.1 8B & 0.405 & 0.738 & 0.294 & 0.453 & 0.405 & 0.240 & 0.388 & 0.388 \\
 & Mistral 7B & 0.405 & 0.762 & 0.294 & 0.453 & 0.405 & 0.240 & 0.321 & 0.314 \\
 & Phi-3.5 Mini & 0.143 & 0.143 & 0.220 & 0.667 & 0.143 & 0.233 & 0.682 & 0.729 \\
 & Qwen2.5 7B & 0.333 & 0.667 & 0.179 & 0.123 & 0.333 & 0.094 & 0.318 & 0.331 \\
 & \textbf{Gemma 2 9B} \checkmark & \textbf{0.452} & \textbf{0.786} & \textbf{0.358} & \textbf{0.459} & \textbf{0.452} & \textbf{0.318} & \textbf{0.316} & \textbf{0.348} \\
\midrule

\end{tabular}%
}
\end{table}

\subsection{Few-Shot Prompting Classification}

To demonstrate the effectiveness of few-shot prompting for smaller language models, the performance of five candidates was evaluated across the EV and Gasoline Engine HIL scenarios. As demonstrated in Table \ref{tab:few_shot_small_observable_windows_grouped}, Gemma 2 9B exhibits remarkably good performance in both scenarios, achieving in the EV case a Top-1 accuracy of 0.962, an F1-score of 0.970, a precision of 0.972, a recall of 0.970 and a Matthews correlation coefficient of 0.940, together with the lowest Brier score of 0.056, thereby confirming both strong discrimination and well-calibrated confidence. The advantage becomes more pronounced in the Gasoline Engine scenario, in which Gemma 2 9B maintains a Top-1 accuracy of 0.833, an F1-score of 0.833, a recall of 0.864 and an MCC of 0.714, whereas the remaining models degenerate towards trivial behaviour, returning identical F1-scores of 0.379, a recall of 0.500, a precision of 0.306 and an MCC of 0.000. This phenomenon can be attributed to the collapse of these models onto a single majority class, which reflects their inability to exploit the provided exemplars under the more challenging gasoline conditions. Furthermore, comparison with Table \ref{tab:zero_shot_small_grouped} reveals that the few-shot small model markedly surpasses every zero-shot medium and large model, since the EV Top-1 accuracy of 0.962 far exceeds the 0.544 ceiling observed without exemplars. Consequently, it can be deduced that the prompting strategy, rather than the parameter count, constitutes the dominant factor governing fault classification quality, and that a carefully selected nine-billion-parameter model guided by few-shot prompting provides an effective and well-calibrated solution for automotive fault diagnosis.

\begin{table}[t]
\centering
\footnotesize
\caption{Performance of small LLMs with the few-shot prompt under various case studies. The best model in each scenario is highlighted in bold and marked with \checkmark.}
\label{tab:few_shot_small_observable_windows_grouped}
\resizebox{\textwidth}{!}{%
\begin{tabular}{llcccccccc}
\toprule
Scenario & Model & Top-1 Acc. & Top-2 Acc. & F1 & Precision & Recall & MCC & Brier $\downarrow$ & ECE $\downarrow$ \\
\midrule
Gasoline Engine & Llama 3.1 8B & 0.500 & 0.812 & 0.442 & 0.478 & 0.484 & 0.422 & 0.354 & 0.312 \\
 & Mistral 7B & 0.562 & 0.708 & 0.487 & 0.675 & 0.571 & 0.503 & 0.271 & 0.271 \\
 & Phi-3.5 Mini & 0.625 & 0.812 & 0.603 & 0.673 & 0.622 & 0.554 & 0.291 & 0.268 \\
 & \textbf{Qwen2.5 7B} \checkmark & \textbf{0.792} & \textbf{0.854} & \textbf{0.768} & \textbf{0.773} & \textbf{0.787} & \textbf{0.744} & \textbf{0.164} & \textbf{0.131} \\
 & Gemma 2 9B & 0.667 & 0.875 & 0.649 & 0.669 & 0.655 & 0.590 & 0.182 & 0.173 \\
\midrule
EV  & Llama 3.1 8B & 0.654 & 1.000 & 0.672 & 0.850 & 0.727 & 0.562 & 0.275 & 0.285 \\
 & Mistral 7B & 0.769 & 0.885 & 0.676 & 0.882 & 0.659 & 0.657 & 0.079 & 0.147 \\
 & Phi-3.5 Mini & 0.885 & 0.885 & 0.760 & 0.929 & 0.750 & 0.825 & 0.093 & 0.125 \\
 & Qwen2.5 7B & 0.923 & 1.000 & 0.860 & 0.944 & 0.833 & 0.879 & 0.076 & 0.192 \\
 & \textbf{Gemma 2 9B} \checkmark & \textbf{0.962} & \textbf{1.000} & \textbf{0.970} & \textbf{0.972} & \textbf{0.970} & \textbf{0.940} & \textbf{0.056} & \textbf{0.137} \\
\midrule
\end{tabular}%
}
\end{table}

Table~\ref{tab:few_shot_big_observable_windows_grouped} summarizes the performance of the medium and LLMs under few-shot prompting. The non-urban scenario is the most challenging, with Top-1 accuracy limited to 0.688--0.729. Gemma~3~27B and Qwen2.5~32B achieve the highest Top-1 and Top-2 accuracies (0.729 and 0.917), while Qwen2.5~32B provides the strongest overall balance, attaining the highest precision (0.782), recall (0.752), MCC (0.683), and the best calibration (Brier score 0.185, ECE 0.077). The increased variability of non-urban driving likely limits classification performance, whereas DeepSeek R1 Distill Qwen~32B and Llama~3.1~70B AWQ INT4 achieve accuracies below 0.708.

In contrast, the highway and urban scenarios exhibit saturated discrimination performance, with several models achieving perfect classification metrics. Consequently, calibration becomes the key differentiator. DeepSeek R1 Distill Qwen~32B and Mistral Small~24B achieve the best calibration in the highway scenario, while Mistral Small~24B and Qwen2.5~32B are highlighted in the urban scenario. Overall, Mistral Small~24B is the only model consistently highlighted across both scenarios, indicating the most robust and generalizable behaviour under few-shot prompting.

\begin{table}[t]
\centering
\caption{Performance of medium and large LLMs with the few-shot prompt across the evaluated HIL driving scenarios. The best model in each scenario is highlighted in bold and marked with \checkmark.}
\label{tab:few_shot_big_observable_windows_grouped}
\resizebox{\textwidth}{!}{%
\begin{tabular}{llcccccccc}
\toprule
Scenario & Model & Top-1 Acc. & Top-2 Acc. & F1 & Precision & Recall & MCC & Brier $\downarrow$ & ECE $\downarrow$ \\
\midrule
Non-Urban scenario & Qwen2.5 14B & 0.688 & 0.896 & 0.674 & 0.753 & 0.711 & 0.632 & 0.243 & 0.187 \\
 & Phi-4 14B & 0.688 & 0.917 & 0.681 & 0.746 & 0.685 & 0.625 & 0.212 & 0.108 \\
 & Mistral Small 24B & 0.708 & 0.854 & 0.683 & 0.754 & 0.721 & 0.661 & 0.203 & 0.153 \\
 & \textbf{Gemma 3 27B} \checkmark & \textbf{0.729} & \textbf{0.917} & \textbf{0.717} & \textbf{0.754} & \textbf{0.728} & \textbf{0.674} & \textbf{0.204} & \textbf{0.123} \\
 & \textbf{Qwen2.5 32B} \checkmark & \textbf{0.729} & \textbf{0.917} & \textbf{0.716} & \textbf{0.782} & \textbf{0.752} & \textbf{0.683} & \textbf{0.185} & \textbf{0.077} \\
 & DeepSeek R1 Distill Qwen 32B & 0.688 & 0.896 & 0.684 & 0.771 & 0.700 & 0.633 & 0.244 & 0.184 \\
 & Llama 3.1 70B AWQ INT4 & 0.708 & 0.896 & 0.680 & 0.712 & 0.706 & 0.649 & 0.200 & 0.075 \\
\midrule
Highway scenario & {Qwen2.5 14B}  & {1.000} & {1.000} & {1.000} & {1.000} & {1.000} & {1.000} & {0.016} & {0.115} \\
 & {Phi-4 14B}  & {1.000} & {1.000} & {1.000} & {1.000} & {1.000} & {1.000} & {0.037} &{0.177} \\
 & \textbf{Mistral Small 24B} \checkmark & \textbf{1.000} & \textbf{1.000} & \textbf{1.000} & \textbf{1.000} & \textbf{1.000} & \textbf{1.000} & \textbf{0.015} & \textbf{0.108} \\
 & Gemma 3 27B & 0.923 & 1.000 & 0.939 & 0.949 & 0.939 & 0.884 & 0.062 & 0.046 \\
 & Qwen2.5 32B & 0.962 & 1.000 & 0.970 & 0.972 & 0.970 & 0.940 & 0.042 & 0.106 \\
 & \textbf{DeepSeek R1 Distill Qwen 32B} \checkmark & \textbf{1.000} & \textbf{1.000} & \textbf{1.000} & \textbf{1.000} & \textbf{1.000} & \textbf{1.000} & \textbf{0.010} & \textbf{0.075} \\
 & Llama 3.1 70B AWQ INT4 & 0.654 & 1.000 & 0.672 & 0.850 & 0.727 & 0.562 & 0.174 & 0.223 \\
\midrule
Urban scenario & Qwen2.5 14B & 0.944 & 1.000 & 0.943 & 0.938 & 0.955 & 0.892 & 0.197 & 0.299 \\
 & {Phi-4 14B}  & {1.000} & {1.000} & {1.000} &{1.000} & {1.000} & {1.000} & {0.138} &{0.367} \\
 & \textbf{Mistral Small 24B} \checkmark & \textbf{1.000} & \textbf{1.000} & \textbf{1.000} & \textbf{1.000} & \textbf{1.000} & \textbf{1.000} & \textbf{0.087} & \textbf{0.286} \\
 & Gemma 3 27B & 0.833 & 1.000 & 0.833 & 0.850 & 0.864 & 0.714 & 0.167 & 0.172 \\
 & \textbf{Qwen2.5 32B} \checkmark & \textbf{1.000} & \textbf{1.000} & \textbf{1.000} & \textbf{1.000} & \textbf{1.000} & \textbf{1.000} & \textbf{0.103} & \textbf{0.314} \\
 & DeepSeek R1 Distill Qwen 32B & 0.889 & 1.000 & 0.887 & 0.889 & 0.909 & 0.798 & 0.147 & 0.200 \\
 & {Llama 3.1 70B AWQ INT4}  & {1.000} & {1.000} & {1.000} & {1.000} & {1.000} & {1.000} & {0.131} & {0.356} \\
\bottomrule
\end{tabular}%
}
\end{table}

\begin{figure*}[t]
    \centering
    \begin{subfigure}[t]{0.49\textwidth}
        \centering
        \includegraphics[width=\linewidth]{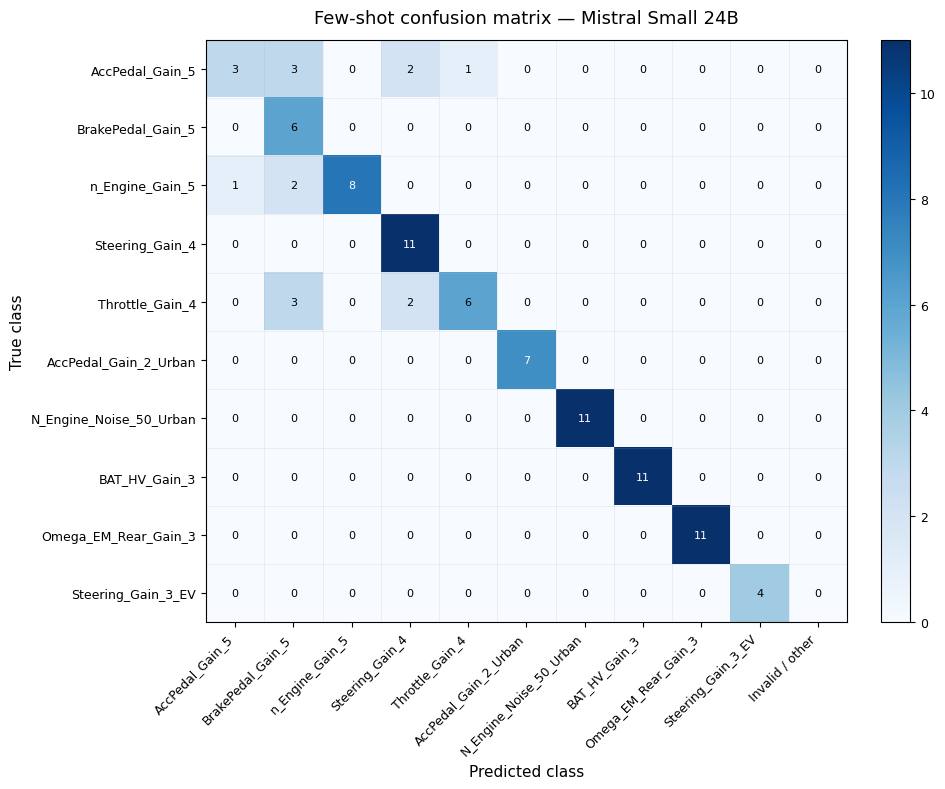}
        \caption{Mistral Small~24B}
        \label{fig:cm_mistral_small_24b}
    \end{subfigure}
    \hfill
    \begin{subfigure}[t]{0.49\textwidth}
        \centering
        \includegraphics[width=\linewidth]{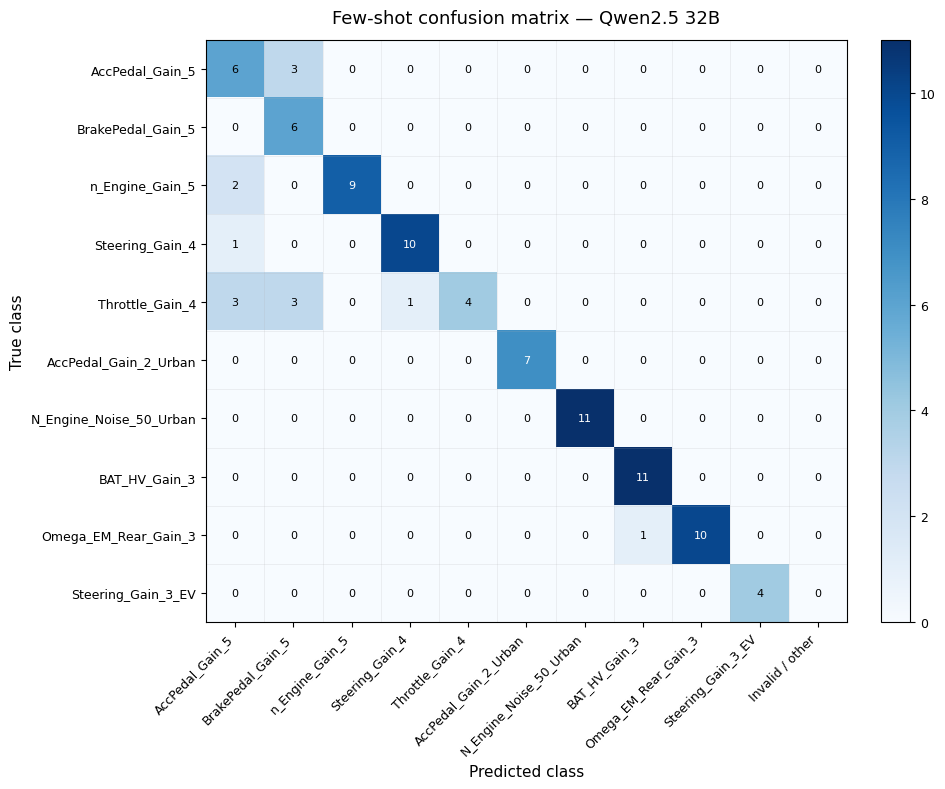}
        \caption{Qwen2.5~32B}
        \label{fig:cm_qwen_32b}
    \end{subfigure}
    \caption{Few-shot confusion matrices for the two strongest models under the few-shot prompting setting. Both models perform strongly overall, while most remaining errors occur among correlated ASM gasoline-engine fault classes.}
    \label{fig:few_shot_confusion_matrices}
\end{figure*}

The confusion matrices in Figure~\ref{fig:few_shot_confusion_matrices} provide a more detailed view of the two strongest models. Both \textit{Mistral Small~24B} and \textit{Qwen2.5~32B} classify the EV and urban-driving faults with high reliability. The clearest errors appear in the ASM gasoline-engine classes, especially among accelerator, brake, throttle, and steering-related faults. This behavior is plausible because these faults can produce correlated downstream effects in vehicle speed, throttle, engine torque, and acceleration. The direct-vs-support prompting mechanism reduces this ambiguity, but it does not remove it completely when several actuator and powertrain signals change together.

Two distinct findings should be separated. Scientifically, few-shot prompting drives classification quality more strongly than parameter count, since Gemma 2 9B with exemplars surpasses every zero-shot medium and large model (Tables 3–4). For deployment, Gemma 2 9B is the most efficient single-model choice where latency and memory are constrained. Within the medium and large tier evaluated under matched conditions in Table 5, Mistral Small 24B is the most robust across all scenarios and is therefore adopted as the reference pipeline model, with Gemma 2 9B recommended as a lightweight alternative.

\subsection{Interpretability of Model Explanations}

In addition to the predicted class, the structured JSON output provides a short explanation of the diagnostic decision. These explanations are useful because they expose whether the model relies on the intended root-signal evidence or mainly on propagated symptoms. Table~\ref{tab:explanations} shows representative correct predictions generated by Mistral Small~24B. The examples cover the EV, ASM gasoline-engine, and ASM urban scenarios.

The examples show that the model not only returns a class label, but also links the decision to the corresponding physical fault location and the relevant signal evidence. For the EV battery fault, the explanation refers to direct voltage evidence in \texttt{V\_Bat\_HV[V]} and supporting battery-related signals. For the ASM accelerator-pedal fault, the model identifies \texttt{Pos\_AccPedal[\%]} as the direct evidence and uses propagated support signals only as additional confirmation. In the urban accelerator-pedal case, the explanation correctly distinguishes medium direct evidence in the accelerator pedal from stronger support evidence in the throttle signal. This behavior is important because propagated signals can be misleading if they are interpreted as root causes. The explanation field therefore provides a lightweight interpretability layer that helps engineers assess whether the predicted class is supported by plausible diagnostic evidence.

\begin{table*}[htbp]
\centering
\scriptsize
\caption{Representative explanation examples from Mistral Small 24B under the few-shot observable-window setting.}
\label{tab:explanations}
\begin{tabular}{@{}l l p{2cm} c p{6.2cm}@{}}
\toprule
\textbf{System} & \textbf{True / predicted class} & \textbf{Fault location} & \textbf{Confidence} & \textbf{Model explanation} \\
\midrule
EV & \texttt{BAT\_HV\_Gain\_3} & High-voltage battery & 0.95 &
Strong direct evidence from \texttt{V\_Bat\_HV[V]} with z=4.684, md=6.671, and strong support from \texttt{I\_Bat\_HV[A]}, \texttt{T\_Bat\_HV[degC]}, and \texttt{SOC\_Bat\_HV[\%]} signals. \\
\addlinespace
Gasoline Engine & \texttt{AccPedal\_Gain\_5} & Accelerator pedal & 0.90 &
The direct evidence for \texttt{AccPedal\_Gain\_5} is strong, with a significant change in \texttt{Pos\_AccPedal[\%]}. The support signals also show medium-level changes, reinforcing the diagnosis. \\
\addlinespace
\bottomrule
\end{tabular}
\end{table*}

\subsection{Runtime and Model Complexity}

The practical value of an LLM-based HIL diagnostic pipeline depends on both predictive performance and computational cost. Table~\ref{tab:few_shot_big_observable_resource_complexity} shows a clear accuracy--runtime trade-off among the evaluated models. Qwen2.5~14B, Phi-4~14B, and Llama~3.1~70B AWQ INT4 provide the fastest inference (2.747--2.977~s per window), whereas Qwen2.5~32B achieves the highest average F1 score but requires the longest inference time (6.614~s).  Mistral Small~24B offers the best compromise, achieving the highest average Top-1 accuracy with a moderate inference time of 4.940~s per window.


GPU memory usage is similar across models (~86 GB), while loading times differ significantly. Since loading is a one-time cost, inference time is more relevant. Overall, Mistral Small 24B offers the best balance between diagnostic performance and computational efficiency, making it the preferred model for the proposed HIL fault-diagnosis pipeline.

\begin{table}[t]
\centering

\caption{Resource and complexity comparison of medium and large LLMs using the observable-window few-shot prompt. Inference-time and token values are averaged under various driving scenarios. Prompt-token counts are computed with each model's native tokenizer. Peak memory denotes the total GPU memory reserved during inference on the evaluation hardware, not the minimal model footprint.}
\label{tab:few_shot_big_observable_resource_complexity}
\resizebox{\textwidth}{!}{%
\begin{tabular}{lccccccc}
\toprule
Model & Inf. time (s) & Throughput & Prompt tokens & Output tokens & Load time (s) & Mem. load (GB) & Peak mem. (GB) \\
\midrule
Qwen2.5 14B & 2.747 & 0.364 & 1502.644 & 99.681 & 100.031 & 86.294 & 86.296 \\
Phi-4 14B & 2.977 & 0.338 & 1502.644 & 122.563 & 98.031 & 86.276 & 86.276 \\
Mistral Small 24B & 4.940 & 0.203 & 1502.644 & 112.817 & 108.034 & 86.353 & 86.353 \\
Gemma 3 27B & 4.773 & 0.210 & 1502.644 & 101.056 & 206.055 & 86.013 & 86.013 \\
Qwen2.5 32B & 6.614 & 0.152 & 1502.644 & 110.087 & 174.051 & 86.241 & 86.243 \\
DeepSeek R1 Distill Qwen 32B & 5.879 & 0.171 & 1502.644 & 92.515 & 82.025 & 86.271 & 86.271 \\
Llama 3.1 70B AWQ INT4 & 2.856 & 0.351 & 1502.644 & 68.039 & 84.023 & 86.333 & 86.333 \\
\bottomrule
\end{tabular}%
}
\end{table}

\section{Conclusion}

The validation of modern automotive software systems produces large, heterogeneous
test recordings whose manual, rule-based analysis is effort-intensive and weak at
detecting faults outside predefined rules. This study addressed the detection and
classification of sensor-related faults during real-time HIL validation of ASSs,
with attention to generalisation and interpretability.

A two-phase framework was developed: a denoising autoencoder trained only on
healthy signals flags abnormal behaviour via reconstruction-error analysis,
removing the need for fault labels, after which the abnormal windows are converted
into compact textual statistical evidence relative to a time-aligned healthy
reference and classified by a frozen large language model through zero-shot and
few-shot prompting, returning the predicted class, ranked alternatives, a confidence
value, the fault location, and a short explanation as a traceable diagnostic output.

The detector achieved average F1-scores of 0.97 across the gasoline-engine and electric-vehicle powertrains and 0.98 across the urban, non-urban, and highway regimes, with each F1-score above 0.96 and an average mean error below 0.03. Zero-shot prompting proved insufficient (best 0.519 F1-score), whereas few-shot prompting reached perfect discrimination
under stable regimes, showing that prompting strategy, rather than parameter count,
governs classification quality, since a nine-billion-parameter model surpassed every
zero-shot medium and large model. Once discrimination saturated, calibration became
decisive, and Mistral Small~24B was adopted as the main pipeline model for its
balance of accuracy, class-balanced reliability, calibration, and inference cost.

The contribution lies in decoupling label-free detection from prompting-based
classification and in an evidence representation separating direct root-signal
deviations from propagated effects, encouraging root-cause reasoning over reliance
on downstream signals. The framework provides interpretable, evidence-linked reports
that let engineers verify a diagnosis on technical grounds, and requires no
task-specific fine-tuning, lowering the cost of adapting to new fault sets.

The evaluation uses real-time HIL simulation data; therefore, performance on real road-test and field data remains to be verified. The taxonomy is limited to ten single-fault classes, and classification relies on leakage-safe few-shot exemplars, as zero-shot performance is considerably lower. Calibration error increases under urban conditions, and the 3--7~s inference time is more suitable for offline than online deployment.

Future work will validate the framework on real-world data, extend it to simultaneous and interacting faults, improve calibration under transient conditions, and investigate LoRA, retrieval-augmented prompting, automated exemplar selection, model distillation, and quantisation.





\bibliographystyle{elsarticle-num-names}
\bibliography{sample}







\end{document}